\documentclass[aps,prd,twocolumn,showpacs,preprintnumbers,amsmath,amssymb,showpacs,showkeys]{revtex4}
\usepackage{graphicx}% Include figure files
\usepackage{dcolumn}% Align table columns on decimal point
\usepackage{bm}% bold math
\usepackage{amssymb}
\usepackage{indentfirst}
\usepackage{epsfig}
\usepackage{epsf}
\usepackage{graphicx}
\usepackage{slashbox}

\newcommand{\eq}{\begin{eqnarray}}
\newcommand{\en}{\end{eqnarray}}

\begin{document}

%\title{Modos con masa variable como alternativa en modelos AdS / QCD con rompimiento de simetr\'ia quiral}
\title{Modes with variable mass as an alternative in AdS / QCD models with chiral symmetry breaking}

\author{Alfredo Vega and
        Ivan Schmidt}

\affiliation{Departamento de F\'\i sica y Centro Cient\'\i fico
Tecnol\'ogico de Valpara\'\i so, Universidad T\'ecnica Federico
Santa Mar\'\i a, Casilla 110-V, Valpara\'\i so, Chile.}

\date{\today}

\begin{abstract}

%Se considera la posibilidad de incorporar de forma satisfactoria
%rompimiento de simetr\'ia quiral en modelos soft wall con dilat\'on
%cuadr\'atico y m\'etrica AdS, para ello permitimos que la masa de
%los modos escalares que se propagan en el bulk dependa de la
%coordenada holografica z. Esto permite un rompimiento espont\'aneo y
%expl\'icito de la simetr\'ia quiral de forma independiente. Adem\'as mostramos que se debe ser cuidadoso con la elecci\'on de la masa dependiente de z, pues para cierta elecci\'on de par\'ametros el modelo entrega algunos resultados inconsistentes con propiedades bien conocidas de QCD.
  We consider a new possibility of incorporating chiral symmetry breaking in soft wall models with a
quadratic dilaton and AdS metric. In particular, we allow the mass of the scalar modes propagating in the
bulk to have a dependence on the holographical coordinate. In one of the models that we discuss, it is
possible obtain a good meson spectra considering current quark masses, and in the other two models
considered, good results can be obtained at the expense of considering rather large quark masses.
\end{abstract}

\preprint{USM-TH-263}

\maketitle

%\section{Introducci\'on}
\section{Introduction}

%Dentro de la variedad de propiedades hadr\'onicas que pueden ser estudiadas mediante AdS / QCD, encontramos modelos que incorporan los efectos del rompimiento de la simetr\'ia quiral expl\'icitamente en el lagrangiano \cite{Erlich:2005qh, Da Rold:2005zs}. Estos modelos en su variante conocida como hard wall permiten que la simetr\'ia quiral sea rota de forma expl\'icita y espont\'anea de modo independiente, pero no describen de forma correcta el espectro hadr\'onico. Esta \'ultima situaci\'on puede ser mejorada mediante la introducci\'on de un campo dilat\'on, que en la mayor\'ia de los trabajos es cuadr\'atico en la coordenada hologr\'afica \cite{Karch:2006pv}, este dilat\'on permite obtener espectros con comportamiento Regge, pero lamentablemente no permite el rompimiento expl\'icito y espont\'aneo de la simetr\'ia quiral  \cite{Colangelo:2008us}.
Within the range of hadronic properties that can be studied using
AdS / QCD, we find models that incorporate effects of chiral
symmetry breaking explicitly in the lagrangian \cite{Erlich:2005qh,
Da Rold:2005zs}. One kind of such models, known as hard wall, allows
to break chiral symmetry both spontaneously and explicitly in an
independent way, but the hadronic spectra calculated in this case
turns out to be not good. This situation can be improved by the
introduction a dilaton field, which in many articles is considered
quadratic in the holographical coordinate z \cite{Karch:2006pv}. The
obtained spectra has Regge behavior, but unfortunately it is not
possible now to break chiral symmetry explicitly and spontaneously
\cite{Karch:2006pv, Colangelo:2008us}.

%La implementaci\'on de rompimiento de simetr\'ia quiral de modo satisfactorio, y sin destruir el comportamiento Regge del espectro hadr\'onico, es un poblema que ha concitado inter\'es en el \'ultimo tiempo. Ejemplos de esta clase de esfuerzos pueden ser hallados en \cite{Zuo:2009dz, Gherghetta:2009ac, Kwee:2007nq, Sui:2009xe}, y ac\'a se presenta una alternativa distinta a la que se puede hallar en la literatura ligada al tema, pues se considera que la masa de los modos que se propagan en el bulk pueden presentar una dependencia sobre la coordenada hologr\'afica, lo que podr\'ia deberse a que los operadores asociados a estos modos poseen dimensiones an\'omalas \cite{Cherman:2008eh}.
The  satisfactory implementation of chiral symmetry breaking,
without sacrificing the hadronic spectra, is a problem that has
attracted much interest lately. Examples of these kind of efforts
can be found in \cite{Zuo:2009dz, Gherghetta:2009ac, Kwee:2007nq,
Sui:2009xe, Zhang:2010tk}. Here we show a different alternative,
since we consider that the mass for modes propagating inside the
bulk can present a dependence on the holographical coordinate z,
which could be due to the fact that operators associated to these
modes might have an anomalous dimension \cite{Cherman:2008eh, Forkel:2008un, Forkel:2010gu, Vega:2008te}.

%En la literatura es posible hallar referencias a masas dependientes
%de z, como en \cite{Cherman:2008eh}, donde los autores sugieren que
%la dimensi\'on an\'omala de los operadores involucrados podr\'ia
%traducirse en masas dependientes de z para los modos duales a dichos
%operadores. Esta idea ha sido usada con \'exito en
%\cite{Vega:2008te}, donde se considera un modelo hologr\'afico que
%sin incluir expl\'icitamente rompimiento de simetr\'ia quiral,
%reproduce el espectro de masa de hadrones con un n\'umero arbitrario
%de constituyentes incluso en el sector bari\'onico, donde como es
%sabido, el dilat\'on al ser factorizado de la ecuaci\'on es incapaz
%mejorar el espectro obtenido con modelos hard wall. Otras
%referencias a masas variables en el bulk, lo podemos hallar en
%trabajos como \cite{Forkel:2007cm, Forkel:2007zz, dePaula:2009za,
%deTeramond:2008ht} donde la masa cambia para diferenciar al pion de
%otros mesones escalares, o en modelos AdS / QCD que consideran
%rompimiento de simetr\'ia quiral, como por ejemplo en
%\cite{Gherghetta:2009ac}, donde el vev se acopla a algunos modos ,
%haciendo que en las ecuaciones que los describen aparezcan
%t\'erminos que pueden ser considerados como masas efectivas.
It is possible to find references to z dependent masses in the
literature \cite{Cherman:2008eh, Forkel:2008un, Forkel:2010gu, Vega:2008te}, where the authors suggest that the
anomalous dimension of operators can be translated into z dependent
masses for dual modes of these operators. This idea was used
successfully in \cite{Vega:2008te}, where an holographic model
without explicit chiral symmetry breaking was considered, and which
can reproduce the hadronic spectrum for spin 1/2 and 3/2 baryons
with with an arbitrary number of constituents. As is known for spin 1/2 case, a
dilaton field can not improve hard wall models, because this field
is factorized from the equation that gives us the spectra \cite{Kirsch:2006he, Vega:2008te}. Other
work related to mass varying in the bulk can be found in papers such
as \cite{Forkel:2007cm, Forkel:2007zz, dePaula:2009za,
deTeramond:2008ht}, where the mass changes in order to differentiate
between the pion and other scalar mesons, or in AdS / QCD models
that consider chiral symmetry breaking, for example in
\cite{Gherghetta:2009ac}, where a vev is coupled to some modes,
producing an effective z depending mass in the equation associated
to some mesons.

%En este trabajo consideramos un modelo AdS / QCD que toma en cuenta rompimiento de simetr\'ia quiral, en el que la masa de los modos escalares depende de la coordenada hologr\'afica z. Cabe destacar que los modos escalares en estos modelos son duales al operador $q\overline{q}$, que no corresponde a una carga conservada, por lo que dicho operador posee dimensiones an\'omalas. De este modo, mostramos que es posible construir un modelo satisfactorio, que incorpora espont\'anea y expl\'icitamente rompimiento de simetr\'ia quiral, incluyendo masas variables. En el modelo presentado, para cierto conjunto de par\'ametros, se obtiene una masa para el meson escalar mas liviano inferior a la masa del Pion, lo que contradice algunas propiedades de QCD \cite{}, pero este problema, afortunadamente no se presenta para todos los casos considerados, por lo que el modelo presentado constituye una alternativa complementaria a esfuerzos como los desarrollados en \cite{Gherghetta:2009ac, Kwee:2007nq, Sui:2009xe}, donde se intenta mejorar los modelos soft wall, mediante la deformaci\'on del dilaton y/o la m\'etrica usadas.
In this article we consider an AdS / QCD model that takes into
account effects of chiral symmetry breaking, with z dependent scalar
mode masses. Notice that these modes are dual to the $q\overline{q}$
operator, which is not a conserved charge, and therefore this
operator has an anomalous dimension. 
%We show that it is possible to build a model that incorporates both spontaneous and explicit chiral symmetry breaking, and which includes variable masses. 
%Consideramos tres modelos, los que para ciertos par\'ametros presentan serios problemas, como obtener masas inferiores a la masa del pion para el meson escalar mas liviano, lo que contradice propiedades de QCD. Este problema para todos los modelos considerados puede ser corregido al considerar masas de quarks grandes, lo que no resulta satisfactorio, pero en el que llamamos modelo II, es posible obtener resultados satisfactorios con masas de corrientes, lo que muestra que este modelo puede ser considerado como una alternativa complementaria a otros modelos ...
We consider three models called models I, II and III, each one related to the behavior of the vev when $z \rightarrow \infty$. For certain set of
parameter we obtain that the lightest scalar meson  has a mass lower
that of the pion, contradicting well established properties of QCD
\cite{Weingarten:1983uj, Witten:1983ut}. Nevertheless, this problem
is not present for all cases, in fact model II allows us to get good results for the spectra with current quark masses, other models considered here needs quark masses with unreasonable large value. Therefore, the model II can be considered as a
serious complementary alternative, different to the effort developed
in \cite{Gherghetta:2009ac, Kwee:2007nq, Sui:2009xe}, where the
authors try to improve soft wall models by deforming the dilaton
and/or the metrics.

%El trabajo consta de las siguientes partes. La secci\'on II incluye una breve descripci\'on del modelo, donde se exhiben las ecuaciones que describen al vev, mesones escalares, vectores y vectores axiales en el lado AdS. En la secci\'on III centramos la atenci\'on en la obtenci\'on de una masa variable para los modos escalares considerados en el modelo. En IV discutimos la fijaci\'on de los par\'ametros que caracterizan al modelo, para pasar luego a la secci\'on V, donde se discute el espectro mes\'onico obtenido con los par\'ametros fijados en IV. Finalmente, la secci\'on VI est\'a dedicada a las conclusiones del trabajo.
The work consists of the following parts. Section II is a brief
description of the model, where we write down the equations that
describe the vev and the scalar, vector and axial vector mesons in
the AdS side. In III we obtain a variable mass for the scalar modes.
In section IV we discuss how to fix the parameters involved in this
model, in order to obtain in section V the spectra with the
parameters of the previous section. Finally, section VI is dedicated
to expose the conclusions of this work.

%\section{Modelo}
\section{Model}

\begin{table}
\begin{center}
\caption{Field content and dictionary of the model.}
\begin{tabular}{ c c c | c c c | c c c | c c c | c c c }
  \hline
  \hline
  & 4D : \textit{O}(x) & & & 5D : $\Phi(x,z)$ & & & p & & & $\Delta$ & & & $m_{5}^{2} R^{2}$ & \\
  \hline
  & $\overline{q}_{L} \gamma^{\mu} t^{a} q_{L}$ & & & $A_{L \mu}^{a}$ & & & 1 & & & 3 & & & 0 & \\
  & $\overline{q}_{R} \gamma^{\mu} t^{a} q_{R}$ & & & $A_{R \mu}^{a}$ & & & 1 & & & 3 & & & 0 & \\
  & $\overline{q}_{R}^{\alpha} q_{L}^{\beta}$ & & & $\frac{1}{z} X$ & & & 0 & & & $3 + \delta$ & & & $m_{5}^{2}(z) R^{2}$ & \\
  \hline
  \hline
\end{tabular}
\end{center}
\end{table}

%Consideramos la versi\'on m\'as usual de modelos soft wall AdS / QCD usando la notaci\'on utilizada en \cite{Gherghetta:2009ac}, que considera un background de tipo AdS en 5D definido por la m\'etrica
We consider the most usual version of soft wall AdS / QCD models,
with the notation used in \cite{Gherghetta:2009ac}, which takes into
account an 5d AdS background defined by
\begin{equation}
 \label{Metrica}
 d s^{2} = \frac{R^{2}}{z^{2}} (\eta_{\mu \nu} dx^{\mu} dx^{\nu} + dz^{2}),
\end{equation}
%donde R es el radio de curvatura AdS, la m\'etrica de Minkowski $\eta_{\mu \nu} = diag (-1, +1, +1, +1)$ y z corresponde a la coordenada hologr\'afica definida en el rango $0 \leq z < \infty$. En este trabajo consideraremos un dilat\'on cuadr\'atico usual
where R is the AdS radius, the Minkowsky metric is $\eta_{\mu \nu} =
diag (-1, +1, +1, +1)$ and z is a holographical coordinate defined
in $0 \leq z < \infty$. In this paper we consider a usual quadratic
dilaton
\begin{equation}
 \label{Dilaton}
 \phi (z) = \lambda^{2} z^{2}.
\end{equation}
%Para describir rompimiento de simetr\'ia quiral en el sector mes\'onico en el lado AdS 5D, la acci\'on considera campos de gauge $SU(2)_{L} \times SU(2)_{R}$ y un campo escalar X. Dicha acci\'on es
To describe chiral symmetry breaking in the mesonic sector in the 5d
AdS side, the action considers $SU(2)_{L} \times SU(2)_{R}$  gauge fields
and a scalar field X. Such action is given by
%\begin{widetext}
\begin{equation}
%\label{Accion}
 S_{5} = - \int d^{5}x \sqrt{-g} e^{-\phi(z)} Tr \biggl[|DX|^{2} + m_{X}^{2} (z) |X|^{2} \nonumber
\end{equation}
\begin{equation}
\label{Accion}
 + \frac{1}{4 g_{5}^{2}} (F_{L}^{2} + F_{R}^{2})\biggr].
\end{equation}
%Esta acci\'on muestra de forma expl\'icita que la masa a considerar para los modos escalares es dependiente de z, siendo este el rasgo que distingue a este modelo de otros que abordan el problema de construir modelos que exhiban rompimiento de simetr\'ia en formalismos de tipo AdS / QCD.
This action shows explicitly that the scalar modes masses are z
dependent, which is the feature that distinguishes this model
from other AdS / QCD models with chiral
symmetry breaking.

%Aqu\'i $g_{5}^{2} = \frac{12 \pi^{2}}{N_{c}}$, con $N_{c}$ el n\'umero de colores, y los campos $F_{L,R}$ estan definidos por
Here $g_{5}^{2} = \frac{12 \pi^{2}}{N_{c}}$, where $N_{c}$ is the
number of colors, and the fields $F_{L,R}$ are defined by
\begin{equation}
 F_{L,R}^{MN} = \partial^{M} A_{L,R}^{N} - \partial^{N} A_{L,R}^{M} - i [A_{L,R}^{M},A_{L,R}^{N}], \nonumber
\end{equation}
%aqu\'i $A_{L,R}^{MN} = A_{L,R}^{MN} t^{a}$, $Tr[t^{a} t^{b}] = \frac{1}{2} \delta^{ab}$, y la derivada covariante es
here $A_{L,R}^{MN} = A_{L,R}^{MN} t^{a}$, $Tr[t^{a} t^{b}] = \frac{1}{2} \delta^{ab}$, and the covariant derivative is
\begin{equation}
 D^{M}X = \partial^{M} X - i A_{L}^{M} X + i X A_{R}^{M}. \nonumber
\end{equation}

%El campo escalar X, que es dual al operador $q\overline{q}$, tiene un vev dado por
The scalar field X, which is dual to the operator $q\overline{q}$, has a vev given by
\begin{equation}
 X_{0} = \frac{v(z)}{2},  \nonumber
\end{equation}
%que rompe la simetr\'ia quiral.
which produces chiral symmetry breaking.

%En la tabla I aparecen los campos incluidos en el modelo y su relaci\'on con los modos que se propagan en el bulk de acuerdo con el diccionario AdS / CFT, donde hacemos incapi\'e en que el operador $q\overline{q}$ posee una dimensi\'on an\'omala ($\delta$), la que se traduce en una masa dependiente de z para los modos duales a dicho operador, de acuerdo con \cite{Cherman:2008eh, Vega:2008te}.
In Table I the fields included in model are shown and also his relationship with
modes propagating in the bulk,  according to AdS / CFT dictionary.
Notice that the operator $q\overline{q}$  has an
anomalous dimension ($\delta$), which in turn produces a mass that
depends on z for the modes dual to this operator, in agreement with
\cite{Cherman:2008eh, Vega:2008te}.

%A partir de (\ref{Accion}), las ecuaciones que describen tanto al vev, como a mesones escalares, vectores y vectores axiales resultan ser
Starting from (\ref{Accion}), the equations that describe the vev and the scalar, vector and axial vector mesons are
\begin{widetext}
\begin{equation}
 \label{vev}
 - z^{2} \partial_{z}^{2} v(z) + z ( 3 + 2 \lambda^{2} z^{2} ) \partial_{z} v(z) + m_{X}^{2} (z) R^{2} v(z) = 0.
\end{equation}
\begin{equation}
 \label{escalar}
 - \partial_{z}^{2} S_{n} (z) + \biggl( \frac{3}{z} + 2 \lambda^{2} z \biggr) \partial_{z} S_{n} (z) + \frac{m_{X}^{2} (z) R^{2}}{z^{2}} S_{n} (z) = M_{S}^{2} S_{n}(z),
\end{equation}
\begin{equation}
 \label{vector}
 - \partial_{z}^{2} V_{n} (z) + \biggl( \frac{1}{z} + 2 \lambda^{2} z \biggr) \partial_{z} V_{n} (z) = M_{V}^{2} V_{n}(z),
\end{equation}
\begin{equation}
 \label{vector axial}
 - \partial_{z}^{2} A_{n} (z) + \biggl( \frac{1}{z} + 2 \lambda^{2} z \biggr) \partial_{z} A_{n} (z) + \frac{R^{2} g_{5}^{2} v^{2} (z)}{z^{2}} A_{n} (z) = M_{A}^{2} A_{n}(z).
\end{equation}
\end{widetext}

%Antes de discutir la fenomenolo\'ia que se obtiene de estas ecuaciones, es necesario discutir la forma que tendr\'a $m_{X}^{2} (z)$, lo que pasamos a revisar en la siguiente secci\'on.
Before discussing the phenomenology of this model,
it is necessary to know the precise form of $m_{X}^{2} (z)$,  something
we will now consider.

%\section{La obtenci\'on de $m_{X}^{2} (z)$}
\section{Obtaining an expression for $m_{X}^{2} (z)$}

\begin{figure*}[ht]
  \begin{tabular}{c c c}
    \includegraphics[width=2.3 in]{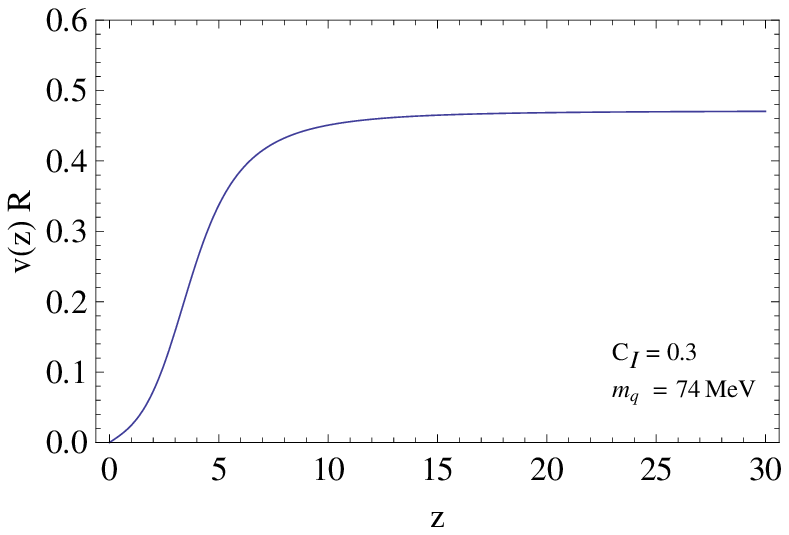}
    \includegraphics[width=2.3 in]{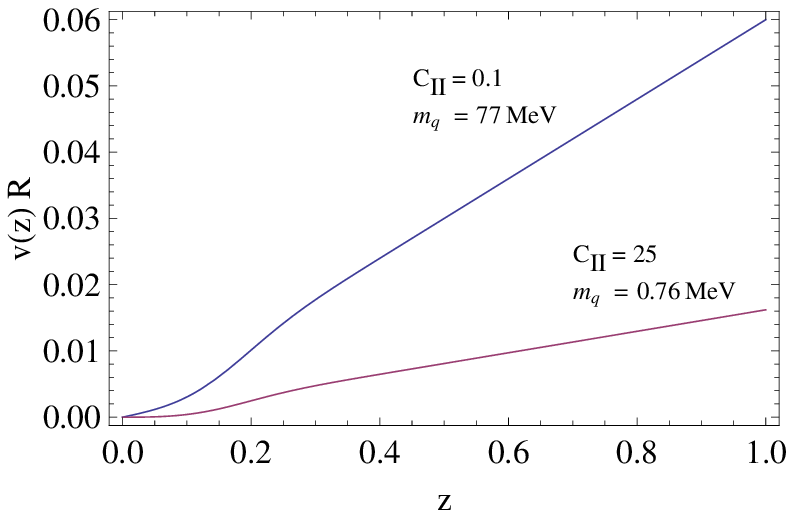}
    \includegraphics[width=2.3 in]{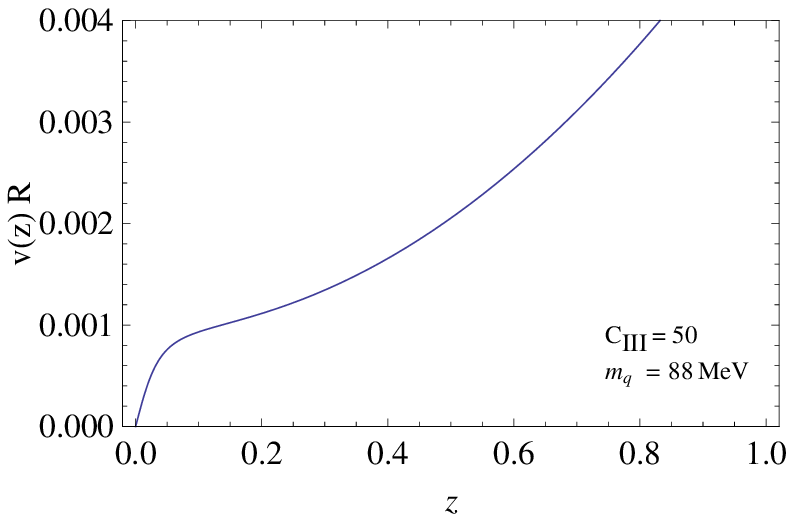} \\
    \includegraphics[width=2.3 in]{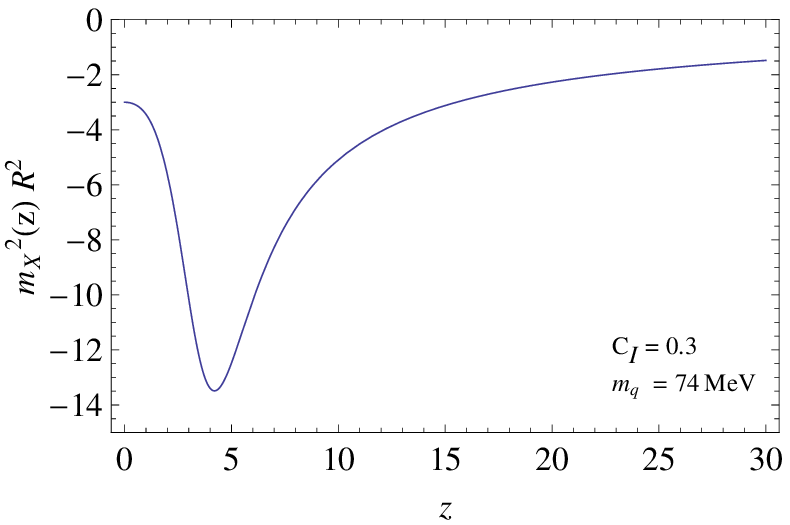}
    \includegraphics[width=2.3 in]{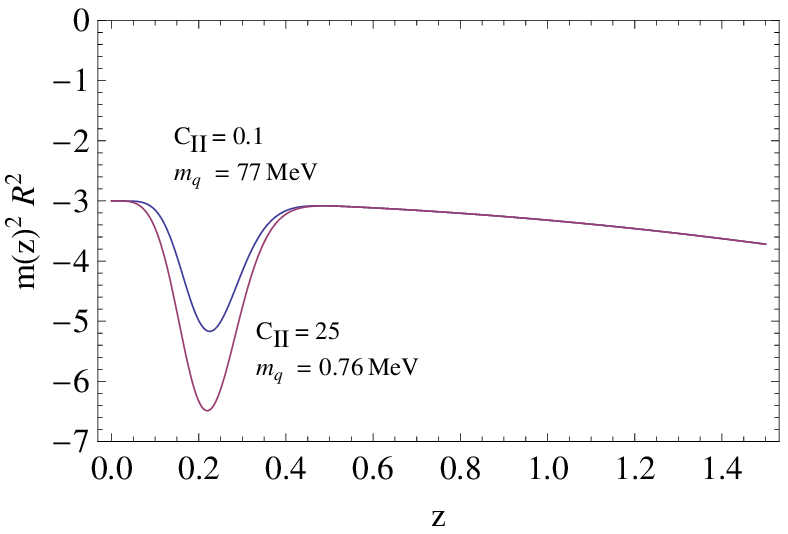}
    \includegraphics[width=2.3 in]{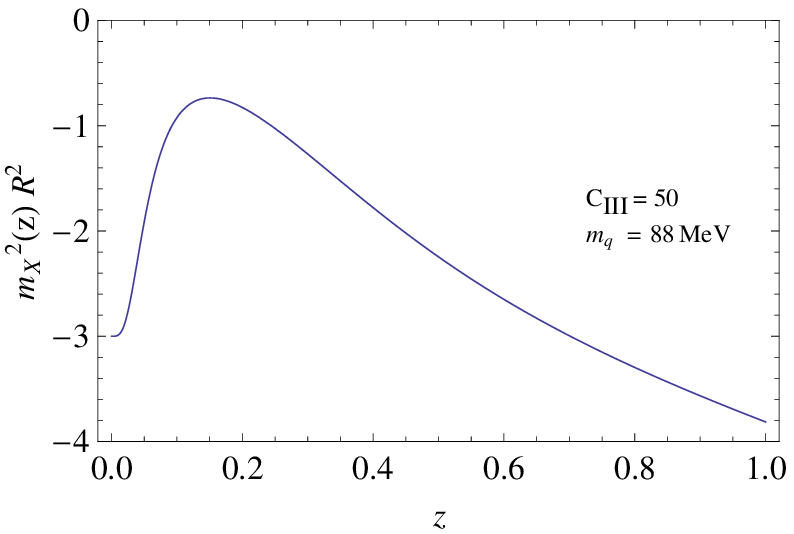}
  \end{tabular}
%  \caption{Los gr\'aficos de arriba muestra la dependencia de $v(z)$ como funci\'on de la coordenada hologr\'afica para distintos valores de $\Omega$, mientras que en los gr\'aficos de abajo podemos ver la variaci\'on respecto de z de la masa de los modos escalares. A izquierda estan los graficos asociados al modelo I, al centro los del modelo II y a la derecha los del modelo III. Para cada gr\'afico se ha usado $\lambda = 0.4$ y el valor de $c_{_{I}}$, $c_{_{II}}$ y $c_{_{III}}$ indicado en cada figura, con lo que se obtienen buenas masas para los mesones.}
\caption{The upper graph shows  $v(z) R$, while in the lower graph
the scalar mode masses as functions of z are given.                                                                                                                      In the left column, we put plots associated to model I, the middle column is model II, and the right column is model III
. All plots have
been obtained with $\lambda = 0.4 GeV$, and $c_{_{I}}$, $c_{_{II}}$ and $c_{_{III}}$ values are
shown in each plot.}
\end{figure*}

%La masa del modo escalar que se propaga en el bulk $m_{X}^{2} (z)$, se obtiene a partir de (\ref{vev}) suponiendo conocida la forma de $v(z)$, que corresponder\'a a una funci\'on capaz de reproducir un par de l\'imites conocidos.
The mass for scalar modes in the bulk,  $m_{X}^{2} (z)$, is obtained starting from (\ref{vev}),  although knowing first the function $v(z)$.  The behavior of this function is known in two limits.

%En primer lugar consideramos el l\'imite usual para $z \rightarrow 0$, seg\'un el cual
First we consider the usual limit $z \rightarrow 0$, according to which
\begin{equation}
 \label{vev Cero}
v(z \rightarrow 0) = \alpha z + \beta z^{3},
\end{equation}
%donde los coeficientes $\alpha$ y $\beta$ est\'an asociados con la masa de los quarks y el condensado quiral respectivamente.
where the $\alpha$ and $\beta$ coefficient are associated with the quark mass and chiral condensate respectively.

%El otro l\'imite considerado para $v(z)$, es $z \rightarrow \infty$. En esta caso exigimos que (\ref{vector axial}) entregue un espectro con comportamiento en el límeite de z grande. Para exponer con mayor claridad esto, transformamos (\ref{vector axial}) usando
The other limit in which we know the behavior of $v(z)$ is when $z
\rightarrow \infty$, and therefore we require that (\ref{vector
axial}) gives a Regge-like in this limit. In order to expose this
clearly, we change (\ref{vector axial}) using
\begin{equation}
 \label{TSchroAxial}
A_{n} (z) = \exp \biggl(\frac{1}{2} \int \biggl(\frac{1}{z} + 2 \lambda^{2} z\biggr) dz \biggr) a_{n} (z).
\end{equation}
%que da origen a una ecuaci\'on de Schr\"odinger cuyo potencial es de la forma
This transformation converts our equation into a Schr\"odinger like equation, with a potential
\begin{equation}
 \label{PotencialAxial}
V_{A} (z) = \frac{3}{4 z^{2}} + \lambda^{4} z^{2} + \frac{R^{2} g_{5}^{2} v^{2} (z)}{z^{2}}.
\end{equation}

%Como es sabido, modelos de tipo soft wall deben reproducir espectros tipo Regge cuando $z \rightarrow \infty$, asi es que el potencial anterior en el l\'imite mencionado debe ser de la forma
As is well known, soft wall models must reproduce spectra with Regge behavior when  $z \rightarrow \infty$, so the potential in this limit must look like
\begin{equation}
 \label{PotencialGeneral}
V (z) = a + b z^{2} + \frac{c}{z^{2}},
\end{equation}
%por lo que $v(z \rightarrow \infty)$ puede ser: constante, lineal o cuadr\'atico en z. En este trabajo solamente consideraremos la primera posibilidad, quedando para un futuro trabajo una discusi\'on m\'as general que considera todos los casos reci\'en mencionados.
and therefore $v(z \rightarrow \infty)$ can be: constant, linear or quadratic in z.

%Ahora que conocemos el comportamiento de $v(z)$ tanto cuando $z \rightarrow 0$ como cuando $z \rightarrow \infty$, elegimos un ansatz capaz de reproducir ambos l\'imites,
Knowing  the behavior of $v(z)$ both for $z \rightarrow 0$ as in $z \rightarrow \infty$, we choose an ansatz capable to reproduce both limits for each possibility,
\begin{equation}
 \label{Ansatz vev I}
v_{_{I}} (z) = \frac{c_{_{I}}}{R} \arctan (A z + B z^{3}), ~~~~(Model~I)
\end{equation}
\begin{equation}
 \label{Ansatz vev II}
v_{_{II}} (z) = \frac{z}{R} (A + B \tanh (c_{_{II}} z^{2})), ~~~~(Model~II)
\end{equation}
\begin{equation}
 \label{Ansatz vev III}
v_{_{III}} (z) = \frac{A z + B z^{3}}{R \sqrt{1 + c_{_{III}}^{2} z^{2}}} . ~~~~(Model~III)
\end{equation}

%Una vez conocida una forma para $v(z)$, el uso de esta en (\ref{vev}) permite obtener una expresi\'on para $m_{X}^{2} (z)$. Tanto $v(z)$ como $m_{X} (z)$ aparecen en la figura XXX, donde $\Omega$ es un par\'ametro arbitrario, y los par\'aetros A y B se relacionan con la masa de los quarks y el condensado quiral, como se muestra a continuaci\'on.
Using these forms for $v(z)$  in (\ref{vev}) allows us to get an
expression for $m_{X}^{2} (z)$. Both $v(z)$ and $m_{X} (z)$ are
shown in Fig. 1 considering only parameters that give us good mesonic spectra. The parameters A and B are related to the quark 
mass and chiral condensate, and its values are fixed according to
the discussion in the next section and $c_{_{I}}$, $c_{_{II}}$ and $c_{_{III}}$ are arbitrary.

%\section{Fijaci\'on de los par\'ametros}
\section{Parameter setting}

%El primer par\'ametro en ser fijado es $\lambda$, el que usualmente se fija con datos de espectros hadrónicos, pues dicho par\'ametro se relaciona con la pendiente Regge. En \cite{Gherghetta:2009ac} los autores consideran una pendiente Regge ajustada al espectro de excitaciones radiales con $n \geq 2$, que finalmente fijan en un valor de $0.732 GeV^{2}$. Ac\'a, debido a que el modelo arroja un espectro con comportamiento Regge para mesones vectore, optamos por fijar $\lambda$ usando un valor cercano a la masa del meson vertorial mas liviano, por lo que finalmente hemos escogido $\lambda = 0.400 GeV$, con lo que se obtiene un buen acuerdo con las masas de mesones vectoriales.
The first parameter that we fix is $\lambda$, using data from the
spectrum. Specifically we consider a specific value for the Regge
slope, which in this kind of models with quadratic dilaton is $4
\lambda^{2}$. In \cite{Gherghetta:2009ac} a Regge slope is fixed
through radial excitations with $n \geq 3$, but in our case, since
the model has Regge behavior in the vector meson sector, we use a
value fixed by the lightest vector meson, so finally we choose
$\lambda = 0.400 GeV$ that allows us to obtain correct value masses
for vector mesons.

\begin{table*}[ht]
\begin{center}
\caption{Scalar meson spectra. All masses are in MeV.}
\begin{tabular}{ c c c | c c c | c c c | c c c | c c c | c c c | c c c }
  \hline
  \hline
  & n & & & $f_{0} (Exp)$ & & & $f_{0}$   & & & $f_{0}$    & & & $f_{0}$   & & & $f_{0}$     & & & $f_{0} (Ref. [6])$ \\
  &   & & &               & & & $m_{q} = 74$ & & & $m_{q} = 77$ & & & $m_{q} = 0.8$ & & & $m_{q} = 88$  & & &    \\
  &   & & &               & & & $c_{_{I}} = 0.3$ & & & $c_{_{II}} = 0.1$ & & & $c_{_{II}} = 25$ & & & $c_{_{III}} = 50$  & & &    \\
  \hline
  & 0 & & & $550^{+250}_{-150}$ & & & 552 & & & 583 & & & 485 & & & 486  & & & 799 & \\
  & 1 & & & $980 \pm 10$ & & & 1089 & & & 1075 & & & 903 & & & 951 & & & 1184 & \\
  & 2 & & & $1350 \pm 150$ & & & 1390 & & & 1350 & & & 1208 & & & 1251 & & & 1466 & \\
  & 3 & & & $1505 \pm 6$ & & & 1621 & & & 1574 & & & 1451 & & & 1490 & & & 1699 & \\
  & 4 & & & $1724 \pm 7$ & & & 1816 & & & 1769 & & & 1659 & & & 1695 & & & 1903 & \\
  & 5 & & & $1992 \pm 16$ & & & 1991 & & & 1943 & & & 1844 & & & 1877 & & & 2087 & \\
  & 6 & & & $2103 \pm 8$ & & & 2149 & & & 2103 & & & 2012 & & & 2042 & & & 2257 &\\
  & 7 & & & $2314 \pm 25$ & & & 2296 & & & 2251 & & & 2166 & & & 2195 & & & 2414 &\\
  \hline
  \hline
\end{tabular}
\end{center}
\end{table*}

%Para los restantes par\'ametros del modelo, de (\ref{Ansatz vev}), que corresponde a la expresi\'on usada para describir el vev, podemos ver que los l\'imites UV e IR de esta corresponden a
The remaining parameters can be fixed using (\ref{Ansatz vev I}), (\ref{Ansatz vev II}) and (\ref{Ansatz vev III}), the expression chosen to describe the vev, for model I
\begin{equation}
 \label{Ansatz vev UV}
v (z \rightarrow 0) = \frac{\Omega}{R} A z + \frac{\Omega}{R} \biggl(-\frac{A^{3}}{3} + B \biggr) z^{3} + O(z^{5}),
\end{equation}
\begin{equation}
 \label{Ansatz vev IR}
v (z \rightarrow \infty) = \frac{\Omega \pi}{2 R} + O(z^{-3}).
\end{equation}

%Comparando (\ref{Ansatz vev UV}) con el valor establecido por el diccionario AdS para el vev \cite{Gherghetta:2009ac}
Comparing (\ref{Ansatz vev UV}) with the value established in the AdS / CFT dictionary, with the notation used in \cite{Gherghetta:2009ac}
\begin{equation}
 \label{vev UV}
v (z \rightarrow 0) = \frac{ m_{q} \zeta}{R} z + \frac{\sigma}{R \zeta} z^{3},
\end{equation}
% Ac\'a hemos usado el par\'ametro $\zeta$ introducido \cite{Cherman:2008eh}, que permite una correcta normalizaci\'on, y cuyo valor esta dado por $\zeta = \sqrt{3}/(2 \pi)$. Con esto, los par\'ametros A y B corresponden a
The parameter $\zeta$  was introduced in \cite{Cherman:2008eh} to get the right normalization, and his value is   $\zeta = \sqrt{3}/(2 \pi)$. With this A and B parameters are given by
\begin{equation}
 \label{Parametro AI}
A = \frac{\sqrt{3}}{2 \pi c_{_{I}}} m_{q}
\end{equation}
%y
and
\begin{equation}
 \label{Parametro BI}
B = \frac{2 \pi}{\sqrt{3} c_{_{I}}} \sigma + \frac{\sqrt{3}}{8 \pi^{3} c_{_{I}}^{3}} m^{3}_{q},
\end{equation}
%Donde $m_{q}$ es la masa de los quarks y $\sigma$ es el condensado quiral.
where $m_{q}$ is the quark mass and $\sigma$ is the chiral condensate.

In a similar way, for Model II we get
\begin{equation}
 \label{Parametro AII}
A = \frac{\sqrt{3}}{2 \pi c_{_{II}}} m_{q}
\end{equation}
%y
and
\begin{equation}
 \label{Parametro BII}
B = \frac{2 \pi}{\sqrt{3} c_{_{II}}} \sigma,
\end{equation}
and finally by for Model III we get
\begin{equation}
 \label{Parametro AII}
A = \frac{\sqrt{3}}{2 \pi} m_{q}
\end{equation}
%y
and
\begin{equation}
 \label{Parametro BII}
B = \dfrac{9 \sqrt{3}}{4 \pi} m_{q} + \frac{2 \pi}{\sqrt{3}} \sigma,
\end{equation}

%Para completar la descripci\'on del modelo, es necesario especificar los valores de $m_{q}$ y $\sigma$, pero debido a que ambas se encuentran relacionadas por medio de la relaci\'on GOR $m^{2}_{\pi} f^{2}_{\pi} = 2 m_{q} \sigma$, tan solo es necesario fijar una de ellas. En esta caso, consideramos $m_{\pi} = 139.6 MeV$ y $f_{\pi} = 92.4$ fijamos la masa de los quarks usando
In order to finish the model description, it is necessary to specify the values for $m_{q}$ and $\sigma$, which are related by the GOR $m^{2}_{\pi} f^{2}_{\pi} = 2 m_{q} \sigma$, and therefore we need to fix only one of them. In this case we use $m_{\pi} = 140 MeV$ and $f_{\pi} = 92 MeV$, and we fix the quark mass using
\begin{equation}
 \label{Constante Decaimiento}
f^{2}_{\pi} = - \frac{1}{g^{2}_{5}} \lim_{\epsilon \rightarrow 0} \frac{\partial_{z} A_{0} (0,z)}{z} |_{z = \epsilon},
\end{equation}
%donde $A_{0} (0,z)$ es soluci\'on de (\ref{vector axial}), con $M^{2}_{A} = 0$, y las condiciones de borde usadas son $A_{0}(0,0)=1$ y $\partial_{z} A_{0} (0,z \rightarrow \infty) = 0$.
where $A_{0} (0,z)$ is solution of (\ref{vector axial}), with $M^{2}_{A} = 0$, and the boundary conditions used are $A_{0}(0,0)=1$ and $\partial_{z} A_{0} (0,z \rightarrow \infty) = 0$.

%Si se observa (\ref{vector axial}), la ecuaci\'on para $A_{0} (0,z)$ posee un t\'ermino que depende de $m_{q}$, asi es que el uso de (\ref{Constante Decaimiento}) entrega un $f_{\pi} (m_{q})$, el que aparece graficado en FIG XXX, y muestra que por cada valor de $\Omega$ considerado, hay dos valores posibles para la masa de los quarks, que arrojan una constante de decaimiento para el pion igual a 92 MeV. Para $\Omega = 0.1$ encontramos $m_{q} = 2.8 MeV$ y $m_{q} = 74.3 MeV$; para $\Omega = 0.5$ hallamos $m_{q} = 7.9 MeV$ y $m_{q} = 72.3 MeV$ y cuando usamos $\Omega = 2$ se obtiene $m_{q} = 7.3 MeV$ y $m_{q} = 73.1 MeV$.
As can be observed in (\ref{vector axial}), the $A_{0} (0,z)$
equation as a term that depends on $m_{q}$, so using (\ref{Constante
Decaimiento}) we get $f_{\pi} (m_{q})$. In general we obtain two possible quark masses in each model, one mass can be considered as a current mass, and only Model II allows us to obtain good mesonic spectra.

%\begin{figure}[h]
%    \includegraphics[width=3.0 in]{f.pdf}
%  \caption{El gr\'afico muestra como var'ia la constante de decaimiento del Pion calculada mediante el uso de (\ref{Constante Decaimiento}) como funci\'on de la masa de los quarks. La l\'inea horizontal corresponde al valor $f_{\pi} = 92 MeV$ considerado en este trabajo.}
%\caption{The pion decay constant as function of $m_{q}$, according
%to (\ref{Constante Decaimiento}). The horizontal line corresponds to
%$f_{\pi} = 92 MeV$.}
%\end{figure}

%\section{Espectro mes\'onico}
\section{Mesonic spectrum}

%Ahora que los par\'ametros del modelo han sido fijados, podemos calcular la masa de algunos mesones con el modelo, las que corresponden a los valores propios de (\ref{escalar}), (\ref{vector}) y (\ref{vector axial}). De estas tres ecuaciones, solamente (\ref{vector}) puede ser resuelta anal\'iticamente, por lo que como estrategia general optamos por transformarlas en ecuaciones de Schr\"odinger, para luego resolver num\'ericamente las ecuaciones que no poseen soluciones exactas, usando el programa de MATHEMATICA llamado schroedinger.nb \cite{Lucha:1998xc}, que ha sido adaptado para poder trabajar con los potenciales de las ecuaciones de Schr\"odinger que obtenemos.
Having fixed the parameters of the model, we can calculate masses for some mesons, which correspond to eigenvalues in the equations (\ref{escalar}), (\ref{vector}) and (\ref{vector axial}). In this set of equations, only (\ref{vector}) can be solved analytically. For this reason we prefer to change all equations into Schr\"odinger like ones, and later solve numerically (\ref{escalar}) and (\ref{vector axial}) using a MATHEMATICA code called SCHROEDINGER.nb \cite{Lucha:1998xc},  which was adapted to our potentials.

%\subsection{Mesones escalares}
\subsection{Scalar mesons}

%Aplicando la siguiente transformaci\'on,
Using the following transformation
\begin{equation}
 \label{TSchroEscalar}
S_{n} (z) = \exp \biggl(\frac{1}{2} \int \biggl(\frac{3}{z} + 2 \lambda^{2} z \biggr) dz \biggr) s_{n} (z),
\end{equation}
%la ecuaci\'on (\ref{escalar}) se transforma en una ecuaci\'on de Schr\"odinger cuyo potencial es de la forma
equation (\ref{escalar}) is converted in a Schr\"odinger like equation, with potential given by
\begin{equation}
 \label{Potencial Escalar}
V_{S} (z) = 2 \lambda^{2} + \lambda^{4} z^{2} + \frac{15}{4 z^{2}} + \frac{m_{X}^{2} (z) R^{2}}{z^{2}}.
\end{equation}

%La forma de este potencial exige resolver num\'ericamente la ecuaci\'on de Schr\"odinger asociada, y algunos resultados que se obtienen de la aplicaci\'on del programa de MATHEMATICA usado se exponen en la Tabla II. Como se puede observar en dicha tabla, para $\Omega = 2$, el estado base es menor que la masa del Pion cuando se considera $m_{q} = 7.3 MeV$, lo que contradice un resultado bien establecido de QCD \cite{}. El inconveniente recien se\~nalado no se presenta para el otro valor posible para $m_{q}$. Una situaci\'on similar ocurre al considerar $\Omega = 0.5$, pues al usar $m_{q} = 7.9 MeV$ la masa del estado base sigue siendo inferior a la masa del Pion, aunque la diferencia se ha reducido en comparaci\'on al caso anterior, y finalmente, para el caso cuando se considera $\Omega = 0.1$ el problema no se presenta, aunque la masa del estado base hallada para $m_{q} = 2.8 MeV$ se aleja bastante del valor experimental. En general se observan buenos resultados al considerar el mayor de los $m_{q}$ obtenidos para cada uno de los $\Omega$ considerados.
Since the potential is complicated, we must solve numerically our
Schr\"odinger equation. Some results obtained using a MATHEMATICA
program appear in Table II. As you can see, all models reproduce good spectra, but in general using a higher quark mass, but only Model II offer this possibility with a reasonable quark mass. 
%Notice that when we consider $\Omega = 2$ the ground state mass is smaller than the pion mass, for the case $ m_{q} = 7.3 $ MeV. This contradicts a well-established QCD result \cite{Weingarten:1983uj, Witten:1983ut}. The problem does not occur for the other possible value for $ m_{q}$ associated to $\Omega = 2$. A similar situation occurs when we take $\Omega = 0.5 $ and $ m_{q}  = 7.9 MeV$, since here the mass of the ground state is still lower than the mass of the pion, although the difference is small compared to the previous case. Nevertheless, the problem is not present when we consider $ m_{q} = 72.3$ MeV. Finally, in the case when $\Omega = 0.1$ the problem does not occur, but the mass of the ground state found for $ m_{q} = 2.8 $ MeV is far from the experimental value. In general, good results were observed when we consider the larger $ m_{q} $ obtained for each $\Omega $ considered.

%\subsection{Mesones Vectoriales}
\subsection{Vector mesons}

%Como se puede observar de (\ref{vector}), el espectro no dependen de $\Omega$.
Notice that in this case equation (\ref{vector}) does not have a
dependence on c parameter.

%En este caso, (\ref{vector}) se convierte en una ecuaci\'on de Schr\"odinger al usar
In this case (\ref{vector}) is converted in a Schr\"odinger equation through
\begin{equation}
 \label{TSchroVector}
V_{n} (z) = \exp \biggl(\frac{1}{2} \int \biggl(\frac{1}{z} + 2 \lambda^{2} z \biggr) dz \biggr) \upsilon_{n} (z),
\end{equation}
%y el potencial que se obtiene es
and the potential is
\begin{equation}
 \label{PotencialVector}
V_{V} (z) = \frac{3}{4 z^{2}} + \lambda^{4} z^{2}.
\end{equation}

%Para este potencial es posible extraer un espectro exacto, que es de la forma
For this potential it is possible to get an exact spectrum, which is
\begin{equation}
 \label{Espectro Vectores}
M_{V}^{2} = 4 \lambda^{2} (n + 1).
\end{equation}

%Algunos ejemplos de masas de mesones vectoriales obtenidas con esta expresi\'on aparecen en la tabla III, donde es posible observar que los resultados son satisfactorios en general.
Some examples of masses for this case are shown in Table III, where as you can see the results are satisfactory in general.

\begin{table}[ht]
\begin{center}
\caption{Vector mesons spectra in MeV.}
\begin{tabular}{ c c c | c c c | c c c | c c c }
  \hline
  \hline
  & n & & & $\rho (Exp)$ & & & $\rho (Model)$ & & & $\rho (Ref.[6])$ & \\
  \hline
  & 0 & & & $775.5 \pm 1$ & & & 800 & & & 475 & \\
  & 1 & & & $1282 \pm 37$ & & & 1131 & & & 1129 & \\
  & 2 & & & $1465 \pm 25$ & & & 1386 & & & 1529 & \\
  & 3 & & & $1720 \pm 20$ & & & 1600 & & & 1674 & \\
  & 4 & & & $1909 \pm 30$ & & & 1789 & & & 1884 & \\
  & 5 & & & $2149 \pm 17$ & & & 1960 & & & 2072 & \\
  & 6 & & & $2265 \pm 40$ & & & 2117 & & & 2243 & \\
  \hline
  \hline
\end{tabular}
\end{center}
\end{table}

\begin{table*}[ht]
\begin{center}
\caption{Axial vector mesons spectra. All masses are in
MeV.}
\begin{tabular}{ c c c | c c c | c c c | c c c | c c c | c c c | c c c }
  \hline
  \hline
  & n & & & $a_{1} (Exp)$ & & & $a_{1}$ & & & $a_{1}$ & & & $a_{1}$ & & & $a_{1}$ & & &  $a_{1} (Ref.[6])$ & \\
  &   & & &               & & & $m_{q} = 74$ & & & $m_{q} = 77$ & & & $m_{q} = 0.8$ & & & $m_{q} = 88$  & & &    \\
  &   & & &               & & & $c_{_{I}} = 0.3$ & & & $c_{_{II}} = 0.1$ & & & $c_{_{II}} = 25$ & & & $c_{_{III}} = 50$  & & &    \\
  \hline
  & 0 & & & $1230 \pm 40$ & & & 867 & & & 1788 & & & 811 & & & 799 & & & 1185 & \\
  & 1 & & & $1647 \pm 22$ & & & 1186 & & & 1959 & & & 1133 & & & 1131 & & & 1591 & \\
  & 2 & & & $1930^{+39}_{-70}$ & & & 1427 & & & 2116 & & & 1384 & & & 1386 & & & 1900 & \\
  & 3 & & & $2096 \pm 122$ & & & 1633 & & & 2262 & & & 1601 & & & 1600 & & & 2101 & \\
  & 4 & & & $2270^{+55}_{-40}$ & & & 1816 & & & 2399 & & & 1789 & & & 1789 & & & 2279 & \\
  \hline
  \hline
\end{tabular}
\end{center}
\end{table*}

%\subsection{Mesones vectores axiales}
\subsection{Axial vector mesons}

%Como fue se\~nalado en la secci\'on III, para transformar a (\ref{vector axial}) en una ecuaci\'on de Schr\"odinger se utiliza la transformaci\'on
As was pointed out in section III, we can transform (\ref{vector axial}) into a Schr\"odinger equation using
\begin{equation}
 \label{TSchroAxial 2.0}
A_{n} (z) = \exp \biggl(\frac{1}{2} \int \biggl(\frac{1}{z} + 2 \lambda^{2} z \biggr) dz \biggr) a_{n} (z),
\end{equation}
%con lo que el potencial obtenido es
and with this, the potential obtained is
\begin{equation}
 \label{PotencialAxial 2.0}
V_{A} (z) = \frac{3}{4 z^{2}} + \lambda^{4} z^{2} + \frac{R^{2} g_{5}^{2} v^{2} (z)}{z^{2}}.
\end{equation}

%En la Tabla IV se han tabulado algunas de las masas obtenidas para este caso, tras resolver num\'ericamente la ecuaci\'on involucrada en este caso.
In Table IV we show some masses for axial vector mesons.

%\section{Conclusiones}
\section{Conclusions}

%Se ha estudiado la posibilidad de incorporar rompimiento de simetr\'ia quiral en modelos soft wall introduciendo una dependencia en la coordenada hologr\'afica para la masa de los modos escalares que se propagan en el bulk, lo que se presenta como un complemento a otros tipos de esfuerzos que intentan conseguir este objetivo modificando al dilat\'on, modificando la m\'etrica y/o introduci\'endo t\'erminos c\'ubicos o cu\'articos en el sector escalar de la acci\'on \cite{Gherghetta:2009ac, Kwee:2007nq, Sui:2009xe}.
The possibility of incorporating chiral symmetry breaking in soft
wall models, introducing a dependence on the holographical
coordinate in the mass for models propagating inside the bulk, was
studied. This idea could be considered as a complement to other
mechanisms that try to solve this problem introducing changes in the
dilaton field, changes in the metric, or introducing a cubic or
quartic term for scalars in the action \cite{Gherghetta:2009ac,
Kwee:2007nq, Sui:2009xe, Zhang:2010tk}.

%En el modelo propuesto, que considera un dilat\'on cuadr\'atico y la m\'etrica AdS usuales, y se ha considerado una expresi\'on para $v(z)$ que es capaz de reproducir el comportamiento esperado en el UV e IR, a partir de la cual es posible obtener una expresi\'on para la masa de los modos escalares en el bulk. Los par\'ametros del modelo han sido fijados usando la relaci\'on GOR, la pendiente Regge y la constante de decaimiento del pion, con lo que se obtienen espectros de masa. Para cierta elecci\'on de par\'ametros se obtiene que la masa del meson escalar mas liviano es inferior a la masa del Pion, lo que contradice un teorema bien establecido de QCD, lo que afortunadamente para el modelo no sucede en todos los casos.
The model considered here considers a usual quadratic dilaton and a AdS
metric, and three expressions  for  $v(z)$ with different limits when z $\rightarrow \infty$ . Starting
with this it is possible to obtain a mass expression for scalar
modes in the bulk. The parameters are fixed using a Gell-Mann-Oakes-Renner relationship
and the pion decay constant. An important point to be considered is that for certain choice of parameters the
mass of the lightest scalar meson is less than the pion mass,
contradicting a well-established QCD result. This can be resolved in
some cases, at the expense of considering rather large quark masses in models I and III. Model II offers the possibility to built a model that reproduces mesonic spectra with a current quark mass, so it is better than the other two.

%A la luz de los resultados expuestos en este trabajo, parece ser que la introducci\'on de masas variables en el bulk debe ser considerada como una alternativa en la construcci\'on de modelos AdS / QCD que consideren rompimiento de simetr\'ia quiral. El espectro obtenido en algunos casos es bastante pobre, por lo que la masa variable del modelo ac\'a presentado no es capaz de dar cuenta de un amplio rango de datos de masas mes\'onicas, pero nos parece que al menos puede ser considerada como un ingrediente adicional en modelos AdS / QCD, pues masas dependientes de z podr\'ian estar asociadas a modos duales a operadores con dimensiones an\'omalas, permitiendo introducir en este tipo de modelos una cantidad que en QCD es importante, pero que con excepci\'on de algunos trabajos, no es considerada al construir modelos AdS / QCD.
We have presented a model that considers AdS modes with variable
mass in the bulk, which certainly requires improvements. Extensions of the model should consider a mixture of ingredients, such as
non quadratic dilatons and / or asymptotically AdS
metrics. Efforts in this direction have already been initiated, and
one of the objectives of this paper is to point that a variable mass
in the bulk can be viewed as a complement to them.

In this paper the variable mass in the bulk is associated with AdS
modes related to operators with anomalous dimensions. In this way a
an important property of QCD was considered, which with the
exception of a few works, is not taken into account when building models of the AdS / QCD type.

%Un aspecto que esta mas all del alcance de este trabajo, pero que merece una menci\n y un posterior estudio, dice relaci\'on con el corrimiento de la masa de los quarks y del condensado quiral. En este trabajo, la introducci\'on de una masa dependiente de z en el bulk esta motivada al considerar la dimensi\'on an\'omala del operador $q\bar{q}$, lo que se deber\'ia traducir en el corrimiento del condensado quiral y de la masa de los quarks. Este aspecto no es considerado en el presente trabajo, donde el \'enfasis ha sido puesto en la posibilidad de obtener rompimiento de simetr\'ia quiral de forma espont\'anea y expl\'icita de forma independiente, y la interpretaci\'on de la masa de los quarks y del condensado quiral se hace en el l\'imite cuando $z \rightarrow 0$, por lo que la dependencia sobre z de dichas cantidades no es considerada en este modelo.
One aspect that is beyond the scope of this work, but it deserves a
mention and further study, is related to the running quark mass and
the chiral condensate. In this work, the introduction of a
z-dependent mass in the bulk is motivated by considering the
anomalous dimension for $q\bar{q}$, which should result in the
running of the chiral condensate and the mass of the quarks.
%, where the emphasis has been placed on the possibility of explicit and spontaneous Chiral Symmetry breaking in an independent way. 
The interpretation of the
mass of the quarks and the chiral condensate is in the limit where $
z \rightarrow 0 $, so the dependence on z of such quantities is not
considered in this model.

%\begin{Agradecimientos}
\begin{acknowledgments}

%Quisi\'eramos agradecer a Dr. Franz Sch\"oberl, quien nos facilit\'o el programa de MATHEMATICA, que con una ligera modificaci\'on fue usado para calcular el espectro mes\'onico. A. V work was supported from Fondecyt (Chile) under Grant No. 3100028.
We would like to thank Ph.D Franz Sch\"oberl, who provided us the MATHEMATICA program that was used with a small change in this work.  Work supported by Fondecyt (Chile) under Grants No. 3100028 and 1100287.

\end{acknowledgments}

\end{document}